\documentclass[11pt]{article}
\usepackage{xspace}
\usepackage{graphicx}
\usepackage{amsmath}
\usepackage{amssymb}
\usepackage{color}

\definecolor{Red}{rgb}{1,0,0}
\definecolor{Green}{rgb}{0,1,0}
\definecolor{Blue}{rgb}{0,0,1}
\definecolor{Black}{rgb}{0,0,0}

\def\beq{\begin{equation}}
\def\eeq#1{\label{#1}\end{equation}}
\def\eeqn{\end{equation}}

\def\beqa{\begin{eqnarray}}
\def\eeqa#1{\label{#1}\end{eqnarray}}
\def\eeqan{\end{eqnarray}}

\let\bar=\overbar

\def\Dslash{\not{\hbox{\kern-4pt $D$}}}
\def\dslash{\not{\hbox{\kern-2pt $\del$}}}

\def\msb{{\bar{\ssstyle M \kern -1pt S}}}

\def\Title#1{\begin{center} {\Large {\bf #1} } \end{center}}

\begin{document}

\Title{Mind the Gap on IceCube: 
Cosmic neutrino spectrum and muon anomalous magnetic moment}

\bigskip\bigskip


\begin{raggedright}  

{\it T. Araki\index{Araki, T.}},
{\it F. Kaneko\index{Kaneko, F.}},
{\it Y. Konishi\index{Konishi, Y.}},
{\it T. Ota\index{Ota, T.},
{\it J. Sato\index{Sato, J.}},
and
{\it T. Shimomura\index{Shimomura, T.}},
\\
Department of Physics,\\
Saitama University,\\
Shimo-Okubo 255, Saitama-Sakura, Japan}\\

\end{raggedright}
\vspace{1.cm}

{\small
\begin{flushleft}
\emph{To appear in the proceedings of the Prospects in Neutrino Physics Conference, 15 -- 17 December, 2014, held at Queen Mary University of London, UK.}
\end{flushleft}
}

\section{Introduction}

In 2013, the IceCube collaboration announced the first discovery 
of two high energy cosmic neutrino events whose energy was around 1
PeV~\cite{Aartsen:2013bka,Aartsen:2013jdh}.
After three years of data taking, they now show the spectrum 
of cosmic neutrino at the energy range between $\mathcal{O}$(100) TeV and 
$\mathcal{O}$(1) PeV~\cite{Aartsen:2014gkd}.
Although the data contain the events originated by atmospheric neutrino,
the hypothesis that all of these events caused by atmospheric neutrino
has already been rejected at more than 5 $\sigma$ confidence level. 
They definitely observe neutrinos that come from astrophysical objects, 
such as active galactic nuclei and gamma-ray burst. 
Although the event number is still small, the spectrum already shows us 
interesting features. 
For example, there is no event observed in the energy range above 3 PeV.
It seems that there is a sharp edge at 3 PeV.
Here, we are motivated by another intriguing feature 
of the spectrum, which is the gap of neutrino events between 400 TeV 
and 1 PeV.
Although the gap has not been statistically established yet, 
it might be an interesting clue of new physics, because 
such a gap structure does not fit to a simple power-law spectrum 
which cosmic ray flux 
often follows.
At the same time,
there is also a long-standing gap in the elementary particle physics,
which is the gap between theory and experiment in 
the muon anomalous magnetic moment.
In this study, 
we try to make a gap in the cosmic neutrino spectrum and fill the gap in
the muon anomalous magnetic moment, introducing one new physics.

\section{IceCube Gap}

The gap in the cosmic neutrino spectrum 
has already been discussed from the particle physics point of view
in many literature.
The relevant new physics falls into the following three categories:
\begin{enumerate}
 \item New physics at source:
       Both of the first two events announced by IceCube had 
       energy of 1 PeV~\cite{Aartsen:2013bka,Aartsen:2013jdh}. 
       This {\it line spectrum} can be explained by 
       two-body decay of a new particle with a mass of 2 PeV,
       and 
       this new heavy particle is a good candidate of dark matter.
       Although the cosmic neutrino spectrum now becomes broad,
       the relation between cosmic neutrino and dark matter,
       which is suggested by this scenario,
       is quite attractive.
       There are recent reanalyses (see e.g., \cite{Rott:2014kfa}).
       
 \item New physics in
       propagation~\cite{Ioka:2014kca,Ng:2014pca,Ibe:2014pja,Blum:2014ewa,Araki:2014ona,Kamada:2015era}:
       Cosmic neutrino with a particular energy 
       gets scattering with the cosmic neutrino background
       through a new interaction.
       Resonant scattering is nice to explain a narrow gap
       in the spectrum.
       
 \item New physics at detection (see e.g., \cite{Barger:2013pla}):
       New charged current interaction in the neutrino 
       detection process can make a bumpy structure in the spectrum.
\end{enumerate} 
We pursue the second possibility, i.e.,
we assume that cosmic neutrinos with the energy corresponding to
the IceCube gap (400 TeV-1 PeV) are scattered by the cosmic neutrino 
background through a new interaction between neutrinos.

\section{Model and muon anomalous magnetic moment}

\begin{figure}[!ht]
\begin{center}
  \includegraphics[width=7cm]{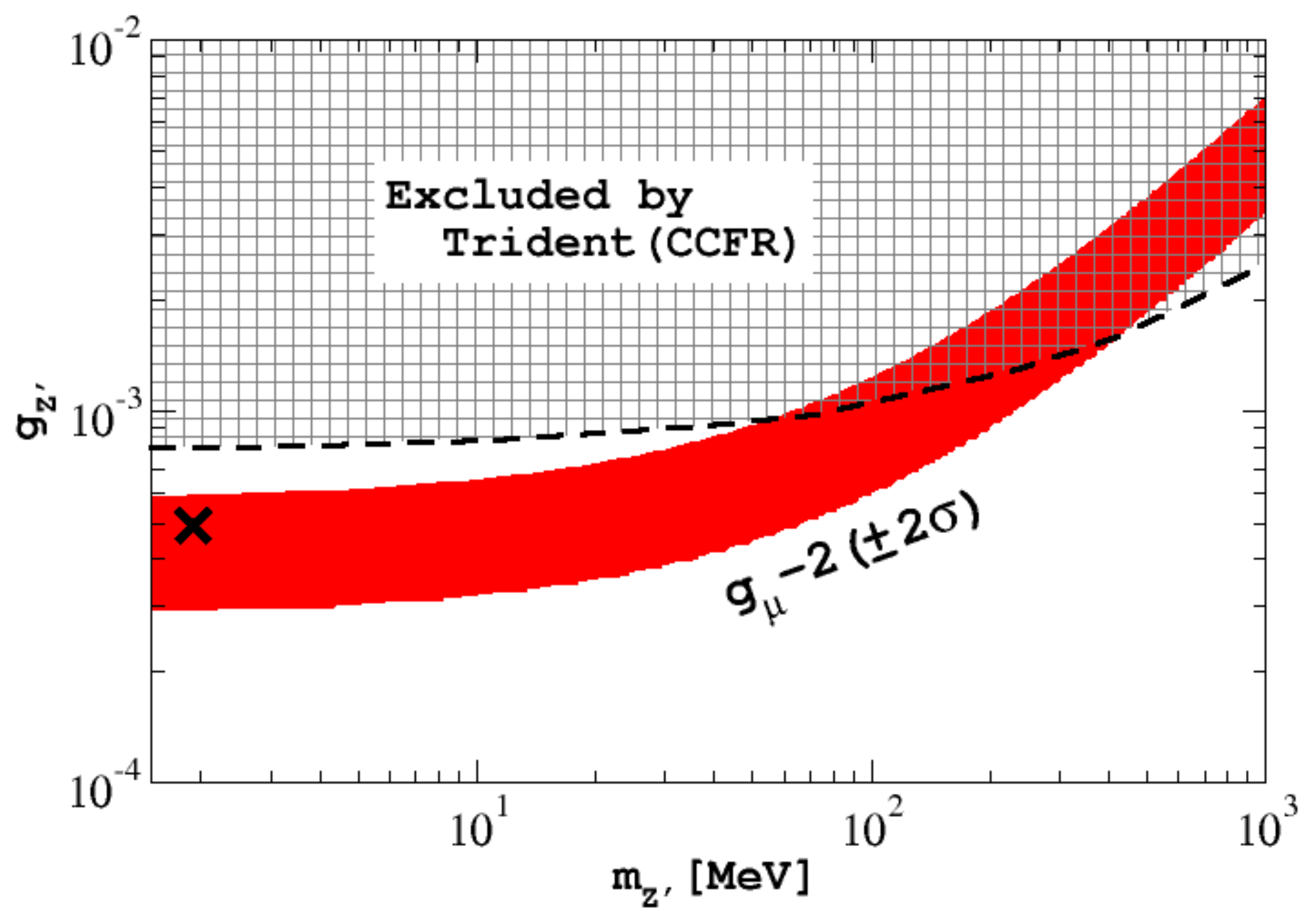}
  \hspace{0.5cm}
  \includegraphics[width=7cm]{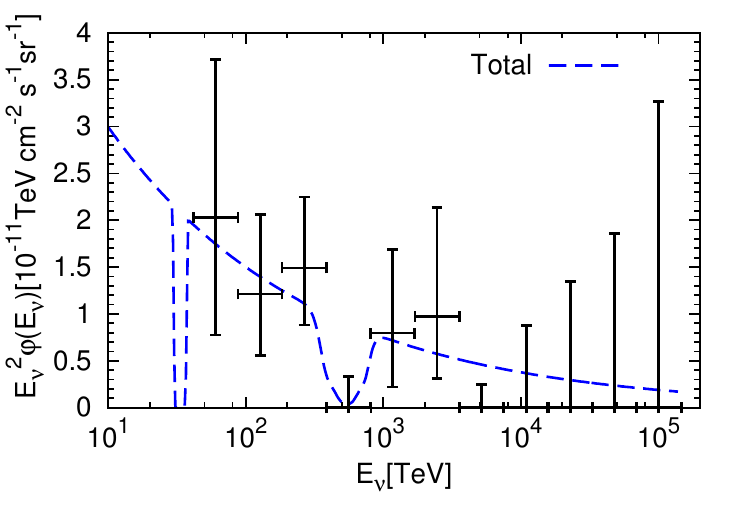}
\caption{[Left] The parameter region favoured 
 by the observation of the muon anomalous magnetic
 moment is indicated by red. The hatched region is excluded by
 the neutrino trident process.
 [Right] The IceCube data is given with crosses.
 The blue dashed curve is a spectrum predicted by 
 the gauged $L_{\mu} - L_{\tau}$ model.
 The plots are taken from \cite{Araki:2014ona}.}
\label{Fig}
\end{center}
\end{figure}

To realise the scenario, we introduce a leptonic gauge interaction
mediated by $Z'$. The introduction of a neutrino interaction 
inevitably brings also a charged lepton interaction through 
the $SU(2)_{L}$ symmetry.
Since a new leptonic interaction with electron is disfavoured 
by a variety of laboratory experiments,
we examine the gauge interaction
associated with muon and tau flavour.
In order to take the gauge anomaly free condition into account,
we assign the opposite charge to tau to 
that of muon~\cite{Foot:1990mn,He:1990pn}. 
The interaction Lagrangians of the model are given as
\begin{align}
 \mathcal{L}_{Z'}
 =
 g_{Z'} \overline{L_{\mu}} \gamma^{\rho} L_{\mu} Z'_{\rho}
 +
 g_{Z'} \overline{\mu_{R}} \gamma^{\rho} \mu_{R} Z'_{\rho}
 -
 g_{Z'} \overline{L_{\tau}} \gamma^{\rho} L_{\tau} Z'_{\rho}
 -
 g_{Z'} \overline{\tau_{R}} \gamma^{\rho} \tau_{R} Z'_{\rho},
\end{align}
which contains not only the neutrino interaction relevant to 
the IceCube spectrum, but also charged lepton interactions.
The charged lepton part, namely $Z'$ interaction with moun, 
gives us a chance to address the 
gap in the muon anomalous magnetic moment.
The left plot in Fig.~\ref{Fig} shows the parameter region
on which $Z'$ makes a contribution to the muon anomalous magnetic 
moment with an appropriate size to explain the observation.
The $Z'$ interaction with charged leptons are constrained by 
various experiments such as colliders and meson decays,
and the most stringent constraint is provided by the measurement 
of the neutrino trident process: 
$\nu_{\mu} N \rightarrow \nu_{\mu} \mu^{-} \mu^{+} X$~\cite{Altmannshofer:2014pba,Mishra:1991bv}.
The region excluded by the trident process is indicated with hatch
on the left plot of Fig.~\ref{Fig}.
The parameter region is narrowed down to the stripe of
$g_{Z'} \sim \mathcal{O}(10^{-4})$ and $M_{Z'} \lesssim 100$ MeV.
We adopt 
\begin{align}
 g_{Z'} = 5.0 \cdot 10^{-4}, \quad M_{Z'} = 2.75 \text{ [MeV]}
\end{align}
as a reference choice of the parameters, which is marked with $\times$
on the plot.

\section{Mean free path and cosmic neutrino flux}

The cross section of 
the scattering process between a cosmic neutrino and 
a cosmic neutrino background (C$\nu$B) is calculated to be
\begin{align}
 \sigma(\nu_{i} \bar{\nu}_{j} \rightarrow \nu \bar{\nu})
 = \frac{g_{Z'}^{2} |g_{ij}|^{2}}{6\pi} 
 \frac{s}{(s-M_{Z'}^{2})^{2} + M_{Z'}^{2} \Gamma_{Z'}^{2}},
\end{align}
where $g_{ij}$ is the $Z'$ coupling in the mass eigenbasis,
$\Gamma_{Z'}$ is the decay width of $Z'$, and
$s$ is the centre of mass energy, which is estimated 
as $2 m_{\nu} E_{\nu}$ at the C$\nu$B rest frame.
In order to make a resonance ($s=M_{Z'}^{2}$)
at the energy $E_{\nu}$ corresponding to the IceCube gap ($\sim 1$ PeV)
with neutrino mass $m_{\nu}$ of $\mathcal{O}(0.1)$ eV, 
the mass of $Z'$ must be set to $\mathcal{O}(1)$ MeV.
We also require that cosmic neutrinos with the energy corresponding to
the IceCube gap do not travel the distance between their sources
and the Earth to reproduce the gap.
The averaged travelling distance of cosmic neutrino can be estimated
with the mean free path $\lambda$ which is roughly given as
$1/(n_{\text{C$\nu$B}} \sigma)$ where $n_{\text{C$\nu$B}}$
is the number density of C$\nu$B in the Universe.
Our requirement, $\lambda \lesssim \mathcal{O}(1)$ Gpc,
leads that the coupling $g_{Z'}$ must be larger than 
$\mathcal{O}(10^{-4})$.
Interestingly, the IceCube gap suggests almost the same parameter
region as the muon anomalous magnetic moment does (cf. Fig.~\ref{Fig}).

Taking account of the effect of C$\nu$B temperature
and the redshift dependence of the mean free path, 
we numerically calculate the cosmic neutrino flux $\varphi(E_{\nu})$
which is shown with the blue dashed curve 
in the right panel of Fig.~\ref{Fig}.
Here we assume that the original cosmic neutrino flux at source
follows a power-law spectrum ($\propto E^{-2.3}$) 
and the source is located at the redshift of $z=0.2$.
For neutrino mass spectrum, we take the inverted hierarchy
and set the lightest neutrino mass to $3.0 \cdot 10^{-3}$ eV.
The curve fits nicely to the observed flux which is indicated
with crosses in the right panel of Fig.~\ref{Fig}.
The details of our setup and calculations can be found in~\cite{Araki:2014ona}.

\section{Summary}

Introducing a new $U(1)_{L_{\mu} - L_{\tau}}$ gauge symmetry,
we have successfully reproduce the gap in the cosmic neutrino spectrum 
reported by the IceCube collaboration, and at the same time 
we have made an additional contribution to
the muon anomalous magnetic moment, which fills the gap between
the standard model prediction and the experimental observation.

\bigskip
\section{Acknowledgments}

This work is supported by Japan society for the promotion of science:
KAKENHI (Grant-in-Aid for scientific research) 
on Innovative area ``Neutrino Frontier'' \#26105503.


\end{document}